\documentstyle[11pt,newpasp,epsf]{article}

\begin{document}

\title{Pulsating Components in Close Binaries}

\author{C.\ Aerts}

\affil{Institute of Astronomy, Department of Physics and Astronomy, Catholic
University of Leuven, Celestijnenlaan 200 B, B-3001 Leuven, Belgium,
conny@ster.kuleuven.ac.be}

\author{P. Harmanec\altaffilmark{1}}

\affil{Astronomical Institute of the Charles University, V. Hole\v sovi\v
ck\'ach 2, CZ-180 00 Praha 8, Czech Republic \& Astronomical Institute of the
Academy of Sciences, CZ-251 65 Ond\v rejov, Czech Republic,
hec@sunstel.asu.cas.cz}

\altaffiltext{1}{Visiting Professor of the Research Council of the Catholic
University of Leuven from Sept. 1 to Nov. 30, 2003; on leave from the Charles
University, Prague, Czech Republic}

\begin{abstract}
We present an overview of pulsating stars in close binaries, focusing
on the question what role the dupliticity plays in triggering and/or
modifying stellar oscillations and on how it can help us to interpret
the oscillatory behaviour of (one of) the components. We give examples
of characteristic types of oscillations observed in binaries: forced
oscillations and free oscillations in both, short- and long-period binaries.
The importance of studies of oscillations in eclipsing binaries is also
pointed out. A list of line-profile and rapid light
variables in close binaries with their basic properties is provided.
No obvious relations among the orbital eccentricity, orbital frequency,
rotational frequency and intrinsic frequencies of oscillations were
found. The value and future prospects of asteroseismic studies of
binary stars are briefly outlined while the complexity of the problem
and its possible complications are also discussed.
\end{abstract}


\section{Introduction}
The study of non-radial stellar oscillations, nowadays called asteroseismology,
has gained much interest after it became clear that their correct
interpretation allows a fine tuning of the models of stellar interior
structure. Such studies have proved very successful
for the Sun and white dwarfs, and more recently also for the
$\delta\,$Scuti, roAp, $\beta\,$Cep and sdB stars. For recent highlights of
the research of stellar oscillations, readers are referred to
Thompson et al.\ 2003 and Kurtz \& Pollard 2004 and references therein.
Basic lecture notes on stellar oscillations can be downloaded from {\tt
http://www.eneas.info}.  Considering that a significant fraction of
oscillating stars are members of multiple systems, it seems legitimate
to ask how much duplicity and atmospheric oscillations affect each other
and what one can learn from studying such relations.

A more general motivation for studying stellar oscillations in
binaries is at least threefold:\\
1. To understand whether duplicity plays an important role
in triggering or modifying the stellar oscillations. This requires systematic
studies of a large sample of oscillators in binaries.\\
2. To study rapid variations in suitably chosen
binaries and to prove that they indeed arise from stellar oscillations and
not from, e.g., corotating structures (see methodological remarks by
Clarke 2003 on this matter). If so, to derive as many
different oscillation frequencies as possible and
to identify their correct pulsational modes. This will allow
to derive the surface stellar rotational
frequency $\Omega$ from the frequency splittings with an incomparably higher
precision than that derived from $v\sin i$ and $i$ estimates.
A comparison with the orbital frequency would then permit to conclude
whether a spin-orbit synchronism had been achieved or not,
{\sl independently of the knowledge of the orientations of the orbital
and rotational planes}.\\
3. The ultimate goal is to apply asteroseismology in all details,
i.e.\ to derive the internal rotation behaviour, accurate stellar ages
and metallicities and to set limits on the amount of convective
core overshooting. This latter quantity is probably the most uncertain
quantity in the present-day stellar interior models while it has
an enormous effect on the course of stellar evolution, especially for
massive stars with large convective cores. The value of the overshooting
parameter has long remained a subject of debate in the literature.
It was estimated from evolutionary isochrone fitting for either some
star clusters or components of eclipsing binaries. These estimates have
invariably led to surprisingly large values from 0.25 to 0.6, expressed
in the units of the pressure scale height. In contrast to it, a recent
independent asteroseismic evaluation of the overshooting parameter
for the very slowly rotating $\beta\,$Cep star HD\,129929 led to a value
below 0.15 (Aerts et al.\ 2003b). This clearly indicates
that the large values of the overshooting parameter found earlier
might stem from rotational mixing which affects the evolution in
a very similar way as the convective overshooting. Clearly,
a detailed asteroseismic study of a slowly rotating pulsator
in an eclipsing binary (or in a cluster) would be the best way of
attacking the difficult problem of the determination of the true value of
the overshooting parameter.

In the following we highlight some illustrative examples of state-of-the-art
studies of oscillations in binaries, deliberately omitting those discussed
in some other contributions presented here (De Cat et al., Freyhammer et al.,
Fremat et al.).  We also do not discuss either oscillations in sdB binaries,
covered by Maxted (these proceedings) or the classical oscillators in binaries
and their role in the PLC relation and the distance determinations.

\section{Search for oscillations in binaries}
There are two basic ways how to search for oscillations in binaries.
One can observe a sample of close binaries and try to discover
short-period variability in (one of) the components. The first such systematic
search, nicknamed SEFONO, was initiated by Harmanec et al.\ (1997).
While SEFONO has led so far to the detection of two binaries with
line-profile variations (Holmgren et al.\ 1997, 1999) it turns out
to be very difficult to prove that the rapid changes are indeed
due to oscillations and not to some other (or additional)
effect having a similar timescale. Moreover, hints of variability
may turn out to be an overinterpretation (see Harmanec et al.\ 1997
vs. Jan\'{\i}k et al.\ 2003).  Yet, the SEFONO approach is the way
to find new oscillating stars among eclipsing binaries.

The other method, to observe a sample of known oscillators and to
discover the binary nature of some of them, turned out to be
more rewarding.  A fine example of the latter strategy is the discovery
of the binary nature of the faint
$\delta\,$Scuti star XX\,Pyx (A4V, V=11.5). While the numerous oscillations of
this star were well established from multisite campaigns (Handler et al.\ 2000),
their seismic modelling had been unsatisfactory (Pamyatnykh et al.\ 1998). The
first step to resolve the discrepancy was made by Arentoft et al.\ (2001) who
discovered a low frequency of 1.73 c/d in the photometric variability, besides
the intrinsic acoustic frequencies in the range 27 to 38 c/d. Such a low
frequency either points towards gravity-mode oscillations or to duplicity. The
latter hypothesis was shown to be the correct one by Aerts et al.\ (2002), who
carried out the first spectroscopic study of the star. XX\,Pyx turns out
to be a circular-orbit binary with a period of 1.$\!\!^{\rm d}$151 and an M3V
companion. A tidal distortion of the primary is the likely cause
of the failure of the earlier seismic modelling, based on the assumption of
a spherical star. De Cat et al.\ (2000) discovered several new spectroscopic
binaries in the sample of southern slowly pulsating B stars selected for
long-term spectroscopic and photometric monitoring by Aerts et al.\ (1999);
see also De Cat et al. (these proceedings) and there are other recent
reports of such discoveries.

\section{Some characteristic case studies}
In the following, we summarize results of several studies which are
typical for various aspects of the problem. A space constraint does not allow
us to review all known short-period variables in close binaries in detail.
However, we include a list of such binaries, currently known to us.

\subsection{Forced oscillations}
Numerous theoretical studies predict the occurrence of forced oscillations in
close binaries due to resonances between dynamical tides and free
gravity-mode oscillations of spherical degree $\ell =2$.
Willems \& Aerts (2002) provide amplitudes and shapes of
the radial-velocity (RV hereafter) curves resulting from
such  oscillations. They find the RV amplitude to increase with increasing
eccentricity. They obtained sinusoidal RV curves whenever the orbital period
was an exact multiple of the one of an $\ell =2$ free oscillation mode
of the star. It is noteworthy, though, that irregular RV curves are predicted
for forced oscillations whenever the orbital period differs slightly
from an exact multiple of the free oscillation period, even for
a point-source companion. It is therefore clear that the periodic changes
in the shapes of the stars, as they pass from periastron to apastron,
must imply very complex variability and perhaps small cyclic changes of
the oscillation periods of tidally induced oscillations near resonance.
Another obvious complication is that the observer detects
oscillations via RV and line-profile variations coming to view
as the star rotates. For eccentric binaries, a spin-orbit synchronization
is impossible, of course, and as the star changes slightly its radius
during its orbital motion, a small cyclic variation in the observed
oscillation frequency becomes inevitable. At the same time, there is no
chance to obtain an accurate value of the frequency separately at periastron
or apastron since there is no reason they should be in phase during
different periastron passages which means that observing the object
for a longer time does not help unless one applies a proper model
in the frequency analyses. A standard Fourier analysis of such
variable signals would recover several close frequencies but this
need not indicate a real multiperiodicity.
One must therefore be very cautious before claiming
observational evidence of intrinsic multiperiodic (forced) oscillations.

A good example of forced oscillations seems to be the slowly pulsating B star
HD\,177863, for which De Cat et al.\ (2000) found an eccentric ($e=0.60$) orbit
with a period of 11.9 days. The dominant intrinsic period in the RV residuals
and in the multicolour photometric variations of the supersynchronous primary
is 1.19 days, i.e.\ exactly 10 times shorter than the orbital period.
Willems \& Aerts (2002) showed that such a frequency is {\it compatible\/}
with a forced oscillation of $\ell=2, m=-2$ for radial orders $n$ between 27
and 53. Additional variability was found for the star by
De Cat \& Aerts (2002) but the significance of these results
needs further verification before an in-depth seismic analysis can be
attempted.

Another remarkable candidate of forced oscillations is the A9/F0V star
HD\,209295. Handler et al.\ (2002) found it to be the primary of an eccentric
($e=0.35$) binary with an orbital period of 3.1 days (frequency of 0.33
c/d). Among the numerous intrinsic frequencies derived by the authors
there is the orbital frequency, low frequencies of 1.13 and 2.30 c/d,
the acoustic-mode $\delta\,$Scuti-type frequency 25.96 c/d, and
their combinations. Handler et al. pointed out the {\it compatibility\/}
of the measured low frequencies with resonantly excited
$\ell=2, m=-2$ modes provided the rotational frequency
of the primary is supersynchronous at 1.85 c/d.

These two cases are currently the best candidates of true
forced oscillations. However, for the reasons outlined above,
they should still not be considered as proven beyond doubt.

\subsection{Free oscillations in short-period binaries}
Smith (1985) proposed that the free acoustic oscillations in
the B1II+B2V binary $\alpha\,$Vir (Spica, $P$=4.$\!\!^{\rm d}$01, $e=0.15$)
are modified by tidal effects but the idea was not developed since then.
Multiple acoustic high-degree modes were recently found for the massive
close binaries $\psi^2\,$Ori (B1II+B2V, $P_{\rm orb}=2.5$\,days, $e=0.05$,
Telting et al.\ 2001) and $\nu\,$Cen (B2IV+?, $P_{\rm orb}=2.6$\,days,
$e=0.0$, Telting \& Schrijvers 2002). Here, the authors found the
oscillations to remain unaltered by the tides. Another remarkable object
in this respect is $\epsilon\,$Per (B0.5IV, $P_{\rm orb}=14.07$\,days,
$e=0.52$, Tarasov et al. 1995), the archetype of line-profile variables
among the early B stars. In spite of its numerous studies,
there is no consensus whether its short-period variability is due to multiple
oscillation modes (Gies et al.\ 1999) or other
complex phenomena such as corotating structures (Harmanec 1999)
or both.  Finally, Lehmann et al. (2001) provided some evidence
that the amplitude of RV oscillations of EN Lac increases near the
periastron passage. All these stars, and a few others (see Table~1)
are situated in the $\beta\,$Cep instability strip
(cf., e.g., Pamyatnykh 1999) and one can expect free non-radial oscillations
to be excited in them via the $\kappa\,$mechanism.
A conservative conclusion is that none of available studies provided
an ultimate proof that free oscillations are altered by the duplicity.

There is no doubt, however, that the studies of oscillations in binaries,
especially the eclipsing ones, must be rewarding. Analyses of RV and light
curves remain to be the most accurate method of determination
of stellar masses, radii, and luminosities, and also effective temperatures,
metallicities and age. Those values then represent an ideal
starting point for the seismic modelling, once oscillation modes have been
found and correctly identified. The range of values in the
parameter space gets substantially limited this way.
This applies also to the {\it overshooting\/} parameter discussed above.
 Unfortunately, only a very limited number of pulsating stars
in eclipsing binaries has been identified so far. We mention two such
objects below:

EN~Lac is a $\beta\,$Cep star in an eclipsing binary. Unfortunately,
only a shallow primary eclipse is observable in the optical region
and no real solution of the light curve is available.
The binary properties are not, therefore, well enough constrained as yet.
The first seismic modelling has been carried out by
Dziembowski \& Jerzykiewicz (1996).
However, their analysis was hampered by the lack of a
unique identification of each of the three detected oscillation modes.
Lehmann et al.\ (2001) recently analyzed a rich set of high-resolution spectra
and refined both the orbital solution and the values of
oscillation frequencies. Aerts et al.\ (2003a) used them to show
that the two largest-amplitude modes are $(\ell,m)=(0,0)$ and (2,0) modes,
respectively.  This identification was then used by Thoul et al.\ (2003a)
in their recent seismic analysis of the star, which led to a tight
mass-metallicity relation for each value of the overshooting parameter.
Clearly, a detection and correct identification of additional frequencies
on the one hand, and setting tighter constraints on binary properties via
observing and solving the orbital light curve on the other hand,
would help enormously to understand the internal structure of the
oscillating primary. An on-going large multisite photometric and
spectroscopic campaign on EN~Lac led by G.\ Handler (Vienna University)
promises such progress for the near future.

Another oscillating star in an eclipsing binary is V539\,Ara (Clausen 1996).
It consists of B3V and B4V pair moving in a slightly eccentric ($e=0.05$)
orbit with a 3.$\!\!^{\rm d}$17 period and exhibiting apsidal motion. Three
candidate gravity modes were detected, with frequencies 0.74, 0.56 and 0.93
c/d. However, until the mode identification will become available, no
seismic modelling is possible.

\subsection{Oscillations in long-period binaries}
The reason why we treat long-period binaries (those
having the orbital periods at least two orders of magnitude longer
than their oscillation periods) separately is that one does
not expect tidal forces would measurably affect their rapid changes.

A textbook example of oscillations in a wide binary is $\alpha\,$Cen~A.
This binary ($P_{\rm orb}=79\,$years, $e=0.52$) is currently the
subject of intense seismic modelling after the detection of 28 solar-like
oscillation modes in the frequency range 1.8 to 2.9 mHz in the G2V primary by
Bouchy \& Carrier (2001) and more recently of 12 such modes in the frequency
range 3 to 4.6 mHz in the K1V secondary by Carrier \& Bourban (2003). The
excitation mechanism and nature of these modes are entirely the same as
for the Sun which permits direct helioseismic applications, this
time in a well-known binary. Earlier seismic studies have already provided
some insights into the internal structure of this binary and led to
a~very accurate age estimate (Bouchy \& Carrier 2002,
Th\'evenin et al.\ 2002, Thoul et al.\ 2003b). However, those studies
were carried out before the discovery of the oscillations of the
secondary. A refinement of the knowledge of internal structure,
based on a simultaneous seismic analysis of both components,
is to be expected soon.

 $\beta\,$Cen is another interesting long-periodic binary, albeit
of an entirely different nature than $\alpha\,$Cen.
Ausseloos et al.\ (2002) have shown $\beta\,$Cen to be an eccentric
($P=357.\!\!^{\rm d}02$, $e=0.814$) binary with two virtually identical
massive components. {\sl Both stars} show clear line-profile variability
while {\sl no } photometric variations above the detection treshold were
found. It is therefore likely that we are dealing here with high-degree
non-radial oscillation modes that tend to cancel out mutually
in photometric data.  Seismic modelling for this star is ongoing.

Guided by the discovery that  $\beta$~Cep is a long-period ($\sim$90~yrs)
binary, Pigulski \& Boratyn (1992) showed that the secular
variation of the main pulsational period of this archetype
is a direct consequence of the light-time effect on the motion in
the binary orbit. If this fact would go unrecognized, the Fourier
analysis would lead to detection of several close frequencies and
a false claim of multiperiodicity.
 Clearly, the claims of multiperiodicity and/or evolutionary period changes
should always be scrutinized from this perspective, even for
seemingly single stars. For example: Harmanec's (1998) showed that
the 1$\!\!^{\rm d}$37 RV period of
the Be star $\omega$~CMa undergoes slow cyclic changes with a possible
period of 5650~d. Subjecting RVs and synthetic data to a standard
Fourier analysis, he recovered three periods:
1$\!\!^{\rm d}$3719, 1$\!\!^{\rm d}$3464 and
1$\!\!^{\rm d}$3536.

A remarkable case
is $\delta$\,Sco with its extreme 0.92 orbital eccentricity and a
nearly 10-yr period. A strong
Balmer emission and a large optical brightening with overlapping
cyclic variations on a time scale of a few months accompanied
the latest periastron passage in 2000. The natural question is
to what extent a steep gradient in the tidal forces is responsible
for these phenomena (see Miroshnichenko et al.\ 2003 for a discussion).
Note that also $\beta$~Cep itself developed Balmer emission during its
recent periastron passage, but no notable light changes.
The ongoing intensive photometric and spectroscopic monitoring of
both stars may shed light on these questions.

\section{Sample of line-profile variables in binaries}

\begin{table}
\caption[]{A list of line-profile variables in binaries. The longitude of
periastron is quoted always for the optical primary, `APS' denotes
a measurable apsidal motion. In column `Type' the usual notations
SB1, SB2, EB, VB, EL, and LT stand  for spectroscopic, eclipsing,
visual, ellipsoidal a light-time orbits.  Because of limited space,
we refer the reader with apologies to the ADS library for the references
on particular objects. See also Fr\'emat et al. for DG~Leo, and
Freyhammer et al. for V381~Car in these proceedings.}
\label{list}
\scriptsize{
\begin{center}
\begin{tabular}{rrrccrclcl}
\hline\noalign{\smallskip}
  Variable       & HD &P$_{\rm short}$&P$_{\rm orb}$& e  &$\omega$&Type&Spectral\\
star name        &number& (days)  & (days) & &($^\circ$)& &class\\
\noalign{\smallskip}\hline\noalign{\smallskip}
  $\vartheta$ Tuc&  3112&0.049-0.053&7.1036  &0.0  & --& SB2&A7IV\\
     $\gamma$ Cas&  5394&0.07-1.487 &203.59  &0.260& 48& SB1&B0.5e+X??\\
    $\varphi$ Per& 10516&0.6        &126.6731&0.0  & --& SB2&B0.5IVe+sdO\\
           RZ Cas& 17138&0.016-0.018&1.195257&0.0  & --& EB &A3V+K0IV\\
        HIP 14871& 19684&0.347-0.366&31.9456 &0.0  & --& SB1&F0IV\\
            X Per& 24534&0.93       &250.3   &0.111&108& LT &O9.5e+X\\
     $\tau^8$ Eri& 24587&0.86423    &459     &0.18 &106& SB1&B5V\\
$\varepsilon$ Per& 24760&0.1598-1.19&14.069  &0.52 &109&SB1&B0.5IV\\
$\vartheta^2$ Tau& 28319&0.037-0.093&140.728 &0.75 & 49&SB1&A7III\\
   $\eta$ Ori Aab& 35411&0.132-0.432&7.989255&0.011&205&EB &B0.5V\\
     $\psi^2$ Ori& 35715&0.093-0.095&2.529   &0.053&APS&SB2&B1II+B2V\\
        V1046 Ori& 37017&0.901      &18.65612&0.468&118& SB2&B2e+B7\\
      $\zeta$ Tau& 37202&0.09-1.6   &132.9735&0.0  & --& SB1&B1e\\
         V696 Mon& 41335&3.6?       &80.860  &0.0  & --& SB1& B2e \\
           XX Pyx&   -- &0.026-0.035&1.151   &0.0  & --& EL & F+M3\\
   $\gamma^2$ Vel& 68273&0.35-1.94  &78.519  &0.38 &261& SB2&WC8+O7.5e\\
           NO Vel& 69144&  ?        &4.82306 &0.0  & --& SB1,EL&B2.5IV\\
           HY Vel& 74560&1.5511     &8.378   &0.24 & 93& SB1&B3IV\\
           GP Vel& 77581& 8.96?     &8.9644  &0.14 &355& EB &B0.5Ibe+X\\
           DG Leo& 85040&           &4.14675 &0.00 & --& SB2&A8IV+A8IV\\
                 &      &0.0818     &70000:  &$>$0 & ? & VB &A8IV\\
         V381 Car& 92024&0.1397-0.1773&8.32457  &0.03 & 65& EB &B1III\\
         V514 Car& 92287&4.6555     &2.90457 &0.0  & --& SB1,EL&B3IV\\
     $\kappa$ Dra&109387&0.54-0.89  &61.5549 &0.0  & --& SB1&B5IIIe\\
      $\beta$ Cru&111123&0.183-0.359&1828.0  &0.38 &293&SB1&B0.5IV\\
     $\alpha$ Vir&116658&0.174      &4.01454 &0.15 &APS&SB2&B1II+B2V\\
           NY Vir& ---  &0.002      &0.101016&0.0: & --&EB &OBVI+M5V:\\
        $\nu$ Cen&120307&0.17       &2.622   &0.00 & --&SB1&B2IV\\
      $\beta$ Cen&122451&0.153-0.503&357.02  &0.814& 63&SB2&B1III\\
         V869 Cen&123515&1.459      &26.036  &0.264&340&SB2&B9IV\\
$\varepsilon$ Lup&136504&0.170-0.177&4.5598  &0.26 &330&SB2&B2IV+B3V\\
           PT Ser&140873&0.86836    &38.927  &0.731&202&SB2&B8III\\
     $\delta$ Sco&143275&0.0975-0.78&3865.   &0.92 & 24&VB+SB1&B0.5IV\\
    $\beta$ Sco A&144217&0.17?      &6.828245&0.30 &APS&SB2  &B0.5IV+B1.5V\\
     $\sigma$ Sco&147165&0.240-0.247&33.012  &0.40 &287&SB1&B1III\\
                 &      &?          &$>$18000&?\\
         V600 Her&149881&0.18       &5.20065 &0.11 &128&SB1&B0.5III \\
         V861 Sco&152667&0.614??    &7.8482  & 0.04&  0& EB &B0Iae+B2V:\\
     $\kappa$ Sco&160578&0.1998     &195.65  &0.488& 93& SB2&B1.5III+B\\
         V539 Ara&161783&1.08--1.78 &3.1691  &0.053&APS&EB  &B3V+B4V\\
        V2502 Oph&167858&0.623-1.307&4.4852  &0.0  & --&SB1 & F1V+M\\
      $\beta$ Lyr&174638&  4.71?    &12.935  &0.0  & --& EB &B8II+B0e::\\
        V4198 Sgr&177863&1.1896     &11.9154 &0.603&182&SB1&B8V\\
        V1765 Cyg&187459& 3 or 8:   &13.37383&0.335&APS&SB2&B0.5Ibe+B0V:\\
           BW Vul&199140&0.20104    &12200.  &0.46 &116&LT&B2III\\
         V832 Cyg&200120&0.07-0.28  &28.1702 &0.20 &271& SB1&B1.5e\\
        V1931 Cyg&200310&0.300-1.20 &146.6   &0.0  & --& SB1&B1e\\
      $\beta$ Cep&205021&0.19048    &33000   &0.65 &194&VB,LT&B1IV\\
        NSV 25820&209295&0.388-0.886&3.10575 &0.352& 31&SB1, EL&F0V\\
         V360 Lac&216200& 1.6738    &10.08541&0.0  & --& SB2&B3e+F9IV\\
           EN Lac&216916&0.169-0.182&12.0969 &0.04 & 65& EB &B2IV\\
           KX And&218393&0.35-4.86  &38.918  &0.0  & --& SB2&B0.5e+K1III\\
           AR Cas&221253&0.510??    &6.0663  &0.24 &APS& SB2&B4V+A6V:\\
\noalign{\smallskip}\hline
\end{tabular}
\end{center}}
\end{table}

In Table\,1 we collected basic information on binaries with rapid
variations known to us at the time of writing. Objects from several
phenomenologically defined groups ($\beta$~Cep, $\delta$~Sct, SPB,
Be, X-ray, WR etc.) are included.
We were unable to find out any clear relations among the various
characteristics of the binary orbits and the rapid frequencies from
this limited sample. Rapid variations are found for stars in circular
as well as highly eccentric orbits. There is some preference for
the values of $\omega$ between 90$^\circ$ and 130$^\circ$. Indeed,
there are 8 such binaries out of 26 with eccentric orbits
while less than 3 would be expected for a uniform distribution.
This is a bit disturbing since a formal solution of RV curves of
Cepheid variables leads to a similar range of $\omega$ values and
one has to keep in mind that -- unless the secondary is directly
observed -- the duplicity of particular objects remains unproven.
This is yet another warning we wish to express here.

\section{Concluding remarks and prospects}
Given the recent progress in seismic modelling of single stars,
we conclude that the prospects of similar studies in close binaries
are very promising though the task is all but simple.
Particularly important is the chance to constrain
the overshooting parameter via asteroseismic analyses of
non-radial oscillations found in eclipsing binaries (or in clusters).
As for single stars, the study of oscillations in binaries requires
huge observational effort and long-term monitoring, even when the individual
oscillation modes have short periods. The beat-periods of the oscillations are
of the order of weeks to years, and aliases have to be excluded from the
frequency analyses. Multisite campaigns are by far the most
efficient way to proceed, but they require considerable organisational
efforts and large international teams of observers. A breakthrough may
be expected as soon as the data from dedicated satellites like MOST or COROT
will become available. However, even then, supporting spectroscopic
campaigns will be needed.

We tried to point out the inherent difficulties involved.
One is to discriminate properly between forced oscillations and
modified or unaffected free oscillations in close binaries.
A way to proceed is to gather numerous multicolour photometry
and high-resolution spectroscopy of an oscillating star in a close binary,
to cover many binary periods.
Other problems are related to possible small cyclic variations
of the oscillation frequencies, due to variable shape of the star
in close eccentric-orbit binaries, and the light-time effect in the
wide ones.
Especially the first situation is a very challenging one. As already
pointed out, even if enough observations are accumulated, the
data cannot simply be splitted into subsets obtained  near peri- and
apastron since the phase of the varying frequency cannot preserved
between two consecutive passages though the same point in the orbit.
A model of a cyclically varying frequency will have to be applied
before the cases of a single cyclically variable frequency and
several close frequencies could safely be distinguished.
One also has to exclude the possibility that the rapid variations
are due to corotating structures in, or slightly above the stellar
photospheres. This is certainly a conceivable possibility
considering that, for instance, several $\beta$~Cep stars were
found to be Be stars and resonances in circumstellar disks are
well established for certain types of binaries.
All these phenomena can lead to variations on time scales
quite comparable to stellar pulsations.

In spite of all obstacles, however, the study of multiple non-radial
oscillations in close binaries is very desirable since it
offers several advantages over similar studies for single stars.
The binary nature is the best way how to constrain tightly
the basic stellar properties, as it was clearly demonstrated during
this meeting on the examples of not only eclipsing but also astrometric
binaries. The number of the latter ones can be expected to increase
dramatically with the current exciting progress in the optical and IR
stellar interferometry.

We end this short review with the hope that at least some readers of it
will be convinced and inspired and will select some of the objects
from our list for their systematic observations and studies.

\section*{Acknowledgements}
This joint study was made possible thanks to the senior fellowship awarded to PH
by the Research Council of the Catholic University of Leuven which allowed his
3-month stay at the Institute of Astronomy in the Department of Physics and
Astronomy. CA is acknowledges financial support from the Fund for Scientific
Research of Flanders through several grants which made it possible to perform
long-term monitoring of pulsating stars during the past decade. The research of
PH was also supported from the research plan J13/98: 113200004 of the Ministry
of Education, Youth and Sports, research plan AV 0Z1 003909 and project K2043105
of the Academy of Sciences of the Czech Republic and from the grant GA~\v{C}R
205/2002/0788 of the Granting Agency of the Czech Republic.
We gratefully acknowledge useful suggestions and a careful proofreading
of this manuscript by Dr.~David~Holmgren.


\begin{references}

\reference Aerts, C., Handler, G., Arentoft, T., et al.\ 2002, MNRAS 333, L35

\reference Aerts, C., Lehmann, H., Briquet, M., et al.\ 2003a, A\&A 399, 639

\reference Aerts, C., Thoul, A., Daszynska, J., et al.\ 2003b, Science 300, 1926

\reference Arentoft T., Sterken C., Handler G.\ 2001, MNRAS 326, 192

\reference Ausseloos, M., Aerts, C., Uytterhoeven, K., et al.\ 2002,
 A\&A 384, 209

\reference Bouchy, F., Carrier, F., 2001, A\&A 374, L5

\reference Bouchy, F., Carrier, F., 2002, A\&A 390, 205

\reference Breger, M., Pamyatnyk, A.A., Zima, W., et al.\ 2002, MNRAS 336, 249

\reference Carrier, F., Bourban, G.\ 2002, A\&A 406, L23

\reference Clarke, D. 2003, A\&A 407, 1029

\reference Clausen, J.V.\ 1996, A\&A 308, 151

\reference De Cat, P., Aerts, C.\ 2002, A\&A 393, 965

\reference De Cat, P., Aerts, C., De Ridder, J., et al.\ 2000, A\&A 355, 1015

\reference Dziembowski, W. A., Jerzykiewicz, M.\ 1996, A\&A 306, 436

\reference Gies, D., Kambe, E., Josephs, T.S., et al., 1999, ApJ 525, 420

\reference Handler, G., Balona, L.A., Shobbrook, R.R.\ 2002, MNRAS 333, 262

\reference Handler G., et al.\ 2000, MNRAS 318, 511

\reference Harmanec, P. 1998, A\&A 334, 558  

\reference Harmanec, P. 1999, A\&A 341, 867  

\reference Harmanec, P., Hadrava, P., Yang, S., et al.\ A\&A 319, 867

\reference Holmgren, D., Hadrava, P., Harmanec, P., et al.\ 1997, A\&A 322, 565

\reference Holmgren, D., Hadrava, P., Harmanec, P., et al.\ 1999, A\&A 345, 855

\reference Jan\'{\i}k, J., Harmanec, P., Lehmann, H., et al.\ A\&A 408, 611

\reference Kennelly, E.J., Walker, G.A.H., Catala, C., et al.\ 1996, A\&A 313,
571

\reference Kurtz, D.W., Pollard, K.\ (eds.), 2004, Variable Stars in the Local
Group, ASP Conf. Ser., in press

\reference Lehmann, H., Harmanec, P., Aerts, C., et al.\ 2001, A\&A 367, 236

\reference Miroshnichenko, A.S., Bjorkman, K.S., Morrison, N.D., et al.\ 2003,
A\&A 408, 305

\reference Pamyatnykh, A.A.\ 1999, Acta Astron. 49, 119

\reference Pamyatnykh A.A., Dziembowski W.A., Handler G., Pikall H.\ 1998, A\&A
333, 141

\reference Pigulski, A., Boratyn, D.A. 1992, A\&A 253, 178 

\reference Schrijvers, C., Telting, J.H.\ 2002, A\&A 394, 603

\reference Smith, M.A.\ 1985, ApJ 297, 224

\reference Tarasov, A.E., Harmanec, P., Horn, J. et al.\ 1995, A\&AS 110, 59

\reference Telting, J.H., Abbott, J.B., Schrijvers, C.\ 2001, A\&A 377, 104

\reference Th\'evenin, F., Provost, J., Morel, P., et al.\ 2002, A\&A 392, L9

\reference Thompson, M.J., Cunha, M.S., Monteiro, M.J.P.F.G. 2003, (eds.)
 Asteroseismology Across the HR Diagram, Astrophys. Space Sci. 284, No. 1

\reference Thoul A., Aerts, C., Dupret, M.-A., et al.\ 2003a, A\&A 406, 287

\reference Thoul, A., Scuflaire, R., Noels, A., et al.\ 2003b, A\&A 402, 293

\reference Willems, B., Aerts, C. 2002, A\&A 384, 441

\end{references}
\end{document}